# Microscale application of column theory for high resolution force and displacement sensing

B.A. Samuel, A.V. Desai, and M.A. Haque

*Department of Mechanical and Nuclear Engineering,*

*The Pennsylvania State University, University Park, PA 16802, USA*

We present the design, fabrication and experimental validation of a novel device that exploits the amplification of displacement and attenuation of structural stiffness in the post-buckling deformation of slender columns to obtain pico-Newton force and nanometer displacement resolution even under an optical microscope. The extremely small size, purely mechanical sensing scheme and vacuum compatibility of the instrument makes it compatible with existing visualization tools of nanotechnology. The instrument has a wide variety of potential applications ranging from electro-mechanical characterization of one dimensional solids to single biological cells.

Displacement and force are two generic mechanical quantities that pervade broad classes of sensors and actuators. Experimental studies in nanoscale science and technology, such as intermolecular interactions, surface and interfacial forces and coupling of thermo-electro-mechanical properties of nanotubes and nanowires call for very high resolution force and displacement sensing. A number of instruments, such as the atomic force microscopes (AFM), optical traps, nano-indenters and micro-electro-mechanical (MEMS) systems [1] have been developed to meet these stringent resolution requirements. Among these, AFM and its derivatives are predominantly used in contemporary nanotechnology research. In this technique, the bending of thin cantilever beams with low stiffness is used to sense very small forces. Force is measured as the product of probe stiffness and tip displacement (usually measured by photo-diodes with angstrom resolution). Commercial probe microscopes can have stiffness as low as 0.01 N/m and hence can resolve force in the pico-Newton range. Even though AFM performance has been continually improved [2, 3], it is practically limited by factors such as precision of fabrication, stiffness calibration, thermo-mechanical noise and structural instability (sample tip interaction, snap-in). High vacuum and ultra-low temperature operation [4, 5] can significantly enhance the force resolution, but at the cost of restricting the experiment to specific environments. Novel approaches in sensing such as the resonant probe [6] can result in very high force resolution, as demonstrated in [2] with 9 nano-Newton for a 500 N/m stiffness probe. Applications of these techniques are also limited to specific operating environments.

MEMS-based force sensors, on the other hand, lag behind probe microscopy because of the superior displacement resolution of the latter [1, 7, 8]. MEMS applications are also environment specific. For example, electrostatic (such as capacitive) sensors cannot perform in electrolytic solutions or in the presence of electromagnetic fields, which makes the sensing/actuating scheme very important for MEMS devices. However, their small size results in the unique advantage of compatibility with various forms of high resolution microscopy for in-situ experimentation [9]. In-situ tests give us not only quantitative and qualitative data simultaneously, but also allow the direct observation of the events during experiments, which becomes increasingly important as the length scale of the test specimens is reduced to the nanoscale. This feature makes MEMS sensors and actuators attractive for nanoscale experimentation.

In this paper, we present a novel force and displacement sensor design that has force resolution comparable to probe microscopy and has very small overall size (1 mm x 1mm). The device can achieve pico-Newton force resolution, even under an optical microscope (1 micron displacement resolution), which compares favorably to probe microscopes where angstrom level displacement sensors are required to achieve the same force resolution. The device operation is purely mechanical in nature and is unaffected by the type of environment (air, vacuum, aqueous, electromagnetic, electrolytic). The extremely small size of the device makes it compatible to almost all the visualization tools in nanotechnology (AFM, SEM, TEM, STM) as well as fluorescence based bio-technology.

The design philosophy exploits post-buckling deformation mechanics of slender columns. A column acts as a very rigid element until a critical buckling load is applied. After that, the column collapses with large lateral displacements for very small incremental loads. As shown in Figure 1(A), the axial deformation of a buckled beam ($\delta$) is amplified by the lateral displacement (D) as,

$$\delta = \pi^2 D^2 /(4L) \quad [10]$$

(1)

For the beam material and geometry shown in Figure 1(B), 1 micron of lateral displacement (easily resolved by optical microscopes) represents about 5 nanometers of axial displacement. While this displacement amplification phenomenon has been utilized before by numerous researchers, the sudden attenuation of beam stiffness (after critical buckling load of 180 μN shown in Figure 1(B)) has so far been viewed as an undesired/failure phenomenon. The axial force-lateral displacement (P-D) relationship given below predicts that structural stiffness as low as $0.0177 \times 10^{-3}$ N/m is achievable. No other structural configuration allows such small stiffness without loss in stability.

$$P = \omega^2[1-(\xi/D)]\kappa + \omega^4 D^2[1-(\xi/D)^3]\kappa/32 \quad [11]$$

(2)

Here κ is the lowest flexural rigidity, ξ is the initial imperfection and $\omega = 2\pi/L$.

In this study, for the first time, we exploit the post-buckling structural stiffness attenuation along with displacement amplification to design a high resolution MEMS force and displacement sensor. A scanning electron micrograph of the sensor is shown in figure 2(A). Here two sets of slender silicon beams (ab/a'b' and cd/c'd') are connected to a center platform, all of which are released from the silicon substrate. The beams ab and a'b' are slightly shorter than the cd and c'd' beams. We use a piezo-motor to apply displacement at the dd' end, while keeping aa' end fixed. Any misalignment of the displacement is neutralized by the auto-alignment beams, which ensure purely axial compressive loads being applied to the slender beams. The cd and c'd' beams buckle first and upon continued displacement, they apply enough load on the beams ab and a'b' to buckle them. After this, the center platform can be assumed to be a rigid body connected in series with two springs with ultra-low stiffness values. Figure 2(B) shows optical micrograph of such a configuration. It is important to note that the symmetric buckling can be ensured by introducing very small initial imperfection (using lithography). Figure 2(C) shows zoomed schematic view of the nano-mechanical testing specimen. One end of this specimen is kept fixed with the substrate while the other end is attached to the movable center platform. The position of the fixed specimen end, with respect to the center platform motion, therefore determines the nature of the applied load (tensile or compressive). At any point in time, the force on the specimen is given by:

$$P_{specimen} = P_{cd} + P_{c'd'} - P_{ab} - P_{a'b'}$$

(3)

The displacement in the specimen is given by the axial displacement of the center platform (or that of the beams ab or a'b'). Therefore, by reading the lateral displacements in the buckled beams and using equations 1 and 2, we can apply and measure pico-Newton level force and nanometer level displacement in a specimen, which makes the device attractive for nano-mechanical testing. Figure 2(D) shows an individual freestanding multiwalled carbon nanotube sandwiched between polymer layers (fabricated using the same processes as the device) that can be used as a nano-mechanical or interfacial test specimen. Figure 2(E) shows a single mouse myeloma cell compressively loaded by the device.

The device operation was simulated using a commercial finite element package ANSYS. Typical device parameters and the resulting force and displacement resolutions are given in Table 1. Devices with chosen parameters were patterned on silicon-on-insulator (SOI) wafer using photo-lithography and subsequently etched with deep reactive ion etching (DRIE). The movable beams and the center platform were released from the substrate using a HF-methanol vapor phase oxide etch. It is important to note that cofabrication of nanoscale specimens (as shown in figure 2(D), which is based on the same fabrication processing steps) with the device is possible. For nano-electro-mechanical testing, the device can be coated with metal films as electrically isolated conductors.

Process uncertainties during lithography and etching may result in significant variation in simulated and actual device performance. To calibrate the device, we cofabricate two guided cantilever beams that can be characterized in the SEM for a very close approximation of their spring constant. We then apply and measure load and corresponding displacement on these beams using the procedure described above. We also simulate the experiment in ANSYS while varying κ, the lowest flexural rigidity. This is because κ is the link between the geometry and the strength of materials. We then tune the κ value to match the simulated buckled configurations (values of the lateral displacements of the sets of beams) with the experimentally observed ones. The κ value corresponding to the best match (shown in figure 3) is taken to be the actual flexural rigidity of the buckling beams.

After calibration, the device was tested with a simple cantilever beam with previously calibrated spring constant of 0.5 N/m. The cantilever was placed perpendicular and out of plane of the device so that it could be loaded at the free end. Experimental data on the lateral displacements of the buckling beams were recorded and compared with ANSYS simulation, this time using the calibrated value of the flexural rigidity and other parameters constant. The experimental and the simulated data (shown in figure 4) matched very well and yielded the same specimen spring constant value. The testing procedure thus corroborates the device calibration scheme as well.

The MEMS-based device developed in this study can apply and measure pico-Newton level force and nanometer level displacement on nano-mechanical test specimens. The experimental setup requires only an optical microscope for raw data recording. The purely mechanical nature of the force and displacement sensing schemes allows the device to be used in vacuum, air, liquid, electromagnetic environment. The extremely small size of the device allows in-situ nano-mechanical testing with the existing electron and probe microscopy techniques and also fluorescence-based microscopy for mechano-biological experiments at the single cell level.

We acknowledge support from the NSF, (CMS- 0411603) for this research. The devices were fabricated at the Penn State Nanofabrication Facility, a member of the NSF's National Nanofabrication Users Network. We also acknowledge Jong Han for his help in DRIE.

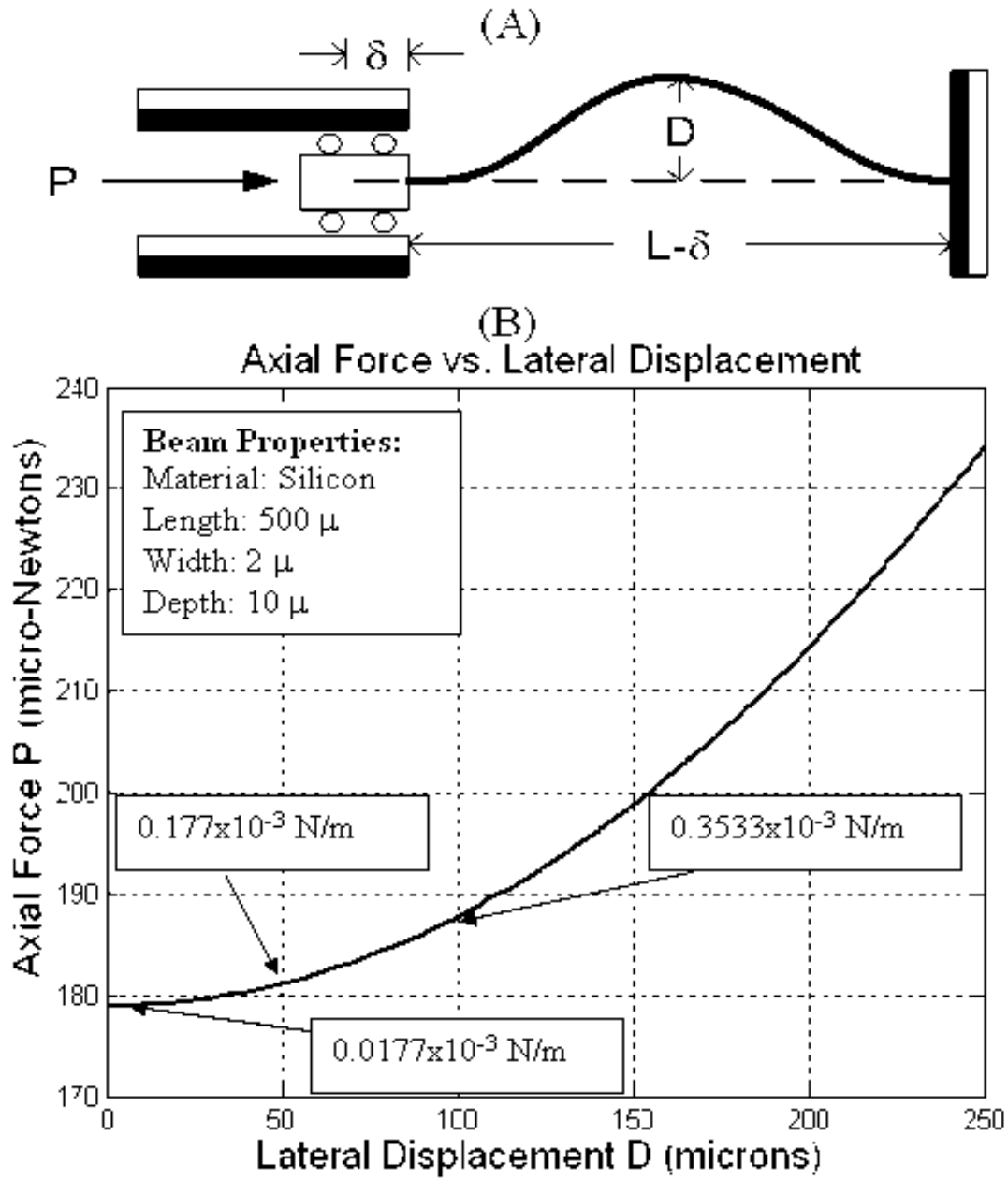

Fig 1. A) Post Buckling Geometry

B) Axial Force vs. Lateral Displacement Plot

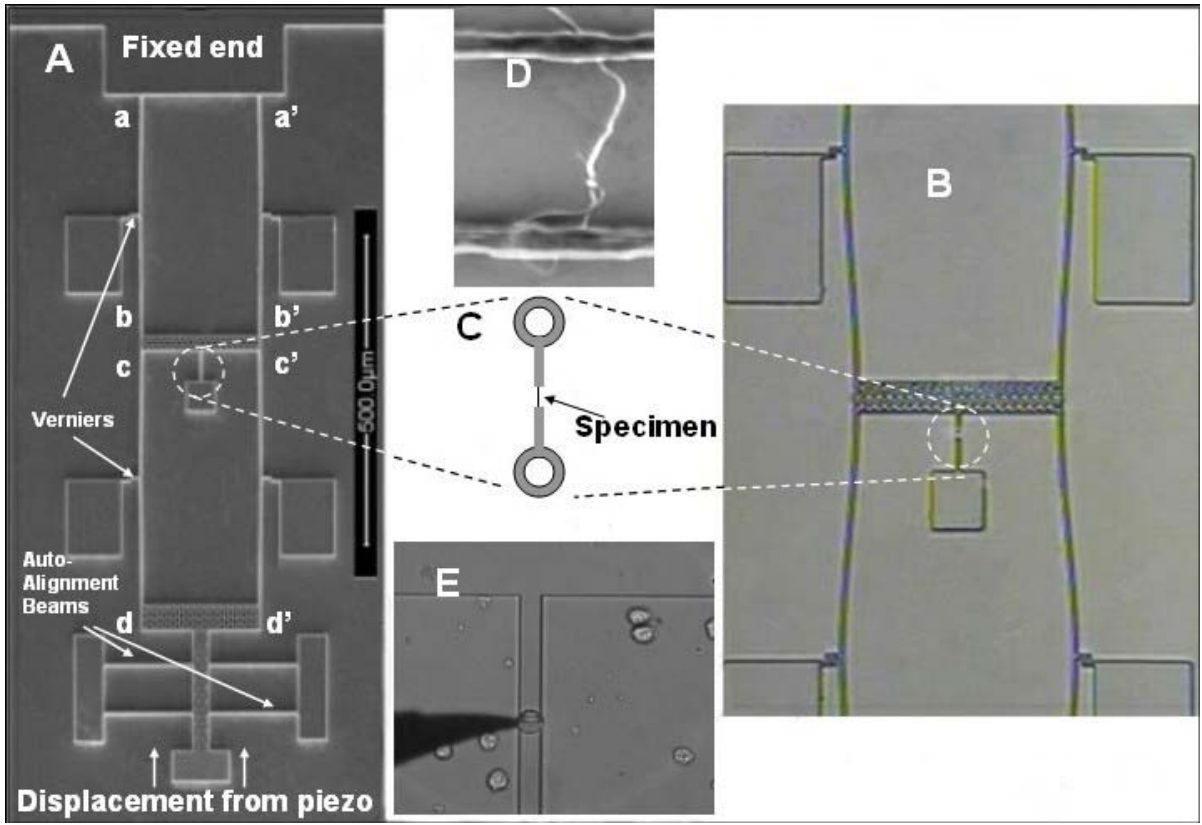

Fig 2.  A) SEM Micrograph of the device
B) Optical Micrograph of Buckled Configuration
C) Specimen fixture schematic
D) Freestanding Carbon Nanotube Specimen
E) Biological Cell Specimen

| Device parameters (microns) | Force on specimen for 1 micron lateral displacement of ab ($D_{ab}$) | Axial displacement ($\delta_{ab}$) for 1 micron displacement of ab ($D_{ab}$) |
|---|---|---|
| ab = 400, cd = 410, bb' = 100, $\xi$ = 5 | 6 pico-Newton | 300 nm |
| ab = 500, cd = 510, bb' = 150, $\xi$ = 3 | 4 pico-Newton | 200 nm |
| ab = 600, cd = 650, bb' = 200, $\xi$ = 2 | 0.5 pico-Newton | 25 nm |

Table 1. Results of FEA Simulations for Varying Parameters

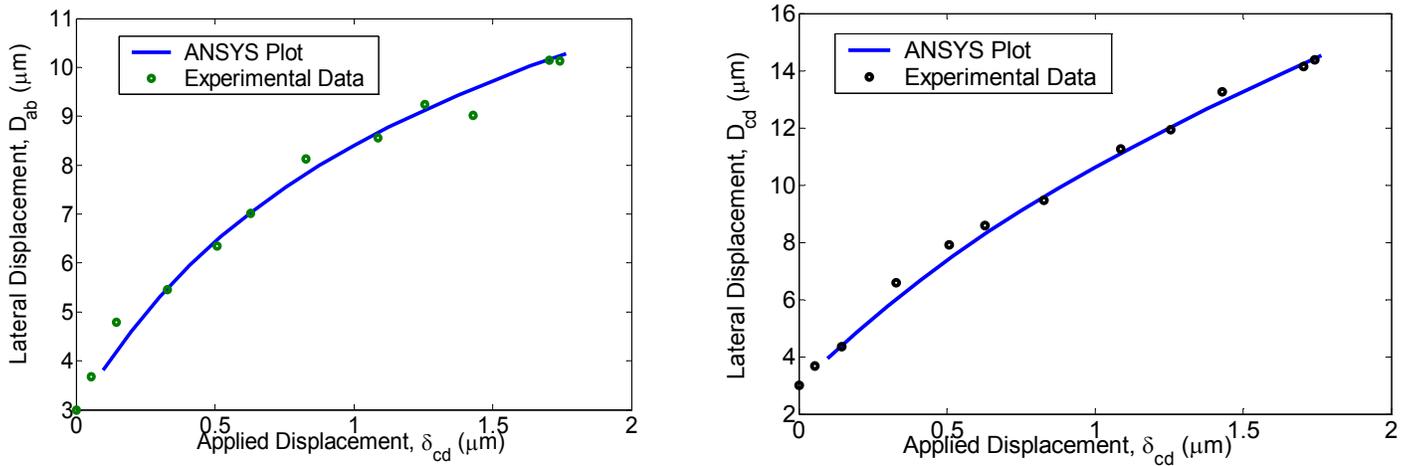

Fig 3. Lateral displacement vs. Applied Displacement Loading – Calibration Plots

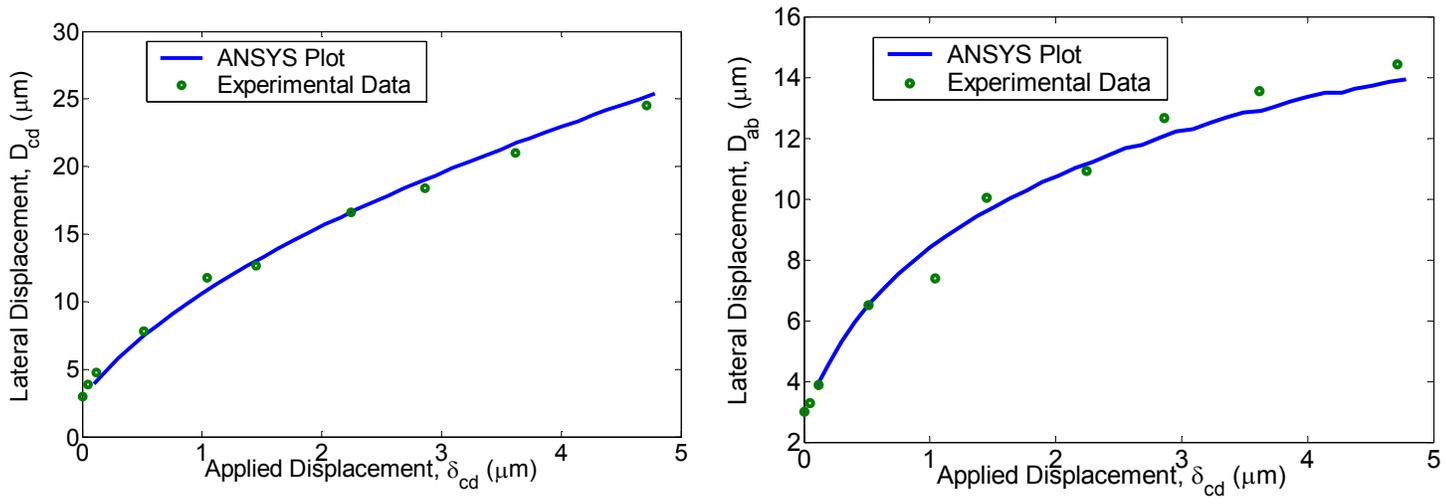

Fig 4. Lateral Displacement vs. Applied Displacement Loading – Experiment with cantilever

acement vs. Applied Displacement Loading – Experiment with
freestanding cantilever